\documentstyle[12pt,draft]{article}

 \newcommand \bi {\bibitem}
 \newcommand \be {\begin{equation}}
\newcommand \bea {\begin{eqnarray} \nonumber }
\newcommand \ee {\end{equation}}
\newcommand \eea {\end{eqnarray}}
 \newcommand \eps {\epsilon}
 \newcommand \si {\sigma}
\newcommand \de {\delta}
\newcommand \De {\Delta}
\newcommand \ga {\gamma}
\newcommand \la {\lambda}

 \newcommand \al {\alpha}
 
 \newcommand \N {{\cal N}}

\newcommand \ba {\overline}

\newcommand \cE  {{\cal E}}

\newcommand \for {\ \ \ \mbox{for}\ \ }

\newcommand {\cp} {\right)}

\def \(  {\left(}
\def \)  {\right(}

\def \form#1 {eq. (\ref{#1}) }
\def \parziale#1#2  {{\partial {#1} \over \partial {#2}}}
 
\begin{document}

\title{Slow dynamics of glassy systems}
\author{  Giorgio Parisi\\
Dipartimento di Fisica, Universit\`a {\em La  Sapienza},\\ 
INFN Sezione di Roma I \\
Piazzale Aldo Moro, Rome 00185}
\maketitle

\begin{abstract}
In these lectures I will present an introduction to the modern way of studying 
the properties of glassy systems.  I will start from soluble models of 
increasing complications, the Random Energy Model, the $p$-spins interacting 
model and I will show how these models can be solved due their mean field 
properties.  Finally, in the last section, I will discuss the difficulties in 
the generalization of these findings to short range models.
\end{abstract}

\section {Introduction}

In these lectures I will study some systems which present some of 
the characteristics of glasses. The aim is to construct simple microscopic models
which can be studied in details and still  behave in interesting way. We will start 
from the simplest ones, where only some of the observed characteristics are 
reproduced and we will later go to more complex systems.

As usual we must select which of the many characteristics of glasses we think 
are important and should be understood.  The main experimental findings about 
glasses that we would like to explain are the following:
\begin{itemize}
\item If we cool the system below some temperature ($T_G$), its energy depends 
on the cooling rate in a significant way. We can visualize $T_G$ as the 
temperature at which the relaxation times become of the order of a hour.

	\item No thermodynamic anomaly is observed: the entropy (extrapolated at 
	ultraslow cooling) is a linear function of the temperature in the region 
	where such an extrapolation is possible.  For finite value of the cooling 
	rate the specific heat is nearly discontinuous.  Data are consistent with 
	the possibility that the true equilibrium value of the specific heat is also 
	discontinuous at a temperature $T_c$ lower than $T_G$.

\item The relaxation time (and quantities related to it, e.g.  the 
viscosity) diverges at low temperature.  In many glasses (the fragile ones) 
the experimental data can be fitted as 
\begin{eqnarray} 
\tau =\tau_0 \exp(\beta B(T))\\ 
B(T) \propto (T-T_c)^{-\la} 
\end{eqnarray}
where $\tau_0 \approx 10^{-13} s$ is a typical microscopic time, $T_c$ is near 
to the value at which we could guess the presence of a discontinuity in the 
specific heat and the exponent $\la$ is of order 1.  The so called Vogel-Fulcher 
law \cite{VF} states that $\la=1$.  The precise value of $\la$ is not well 
determine.  The value $1$ is well consistent with the experimental data, but 
different values are not excluded.
\end{itemize}

These lectures are organized as follows. 

In section II we will study the Random Energy Model (REM) \cite{REM} ; the REM 
may seem to be a too simple model, however many of the main features of the 
model are present in more sophisticated versions, if the appropriate 
modifications are done.

In section III we will study some generalization of the REM in which the 
energies are not fully random, but are partly correlated.  We will see that the 
isolated configurations which dominates the partition function of the REM are now 
promoted to valleys in a corrugated landscape.

Up to this moment we have been in the framework of the mean field theory in 
which all spins interacts among themselves. In section IV I will discuss how 
these results can be hopefully extended to models with short range forces, which 
are the new features and which a the points which are still to be solved.

\section{The Random Energy Model}

\subsection{The definition of the model}

The Random Energy Model \cite{REM} is the simplest model for glassy systems.  It 
have various advantages: it is rather simple (its properties may be well 
understood with intuitive arguments, which may become fully rigorous) and 
display very interesting and new phenomena.

The Random Energy Model is defined as following.  There are $N$ Ising spins 
($\sigma_i$, $i=1,N$) which may take values $\pm1$; the total number of 
configurations is equal to $M\equiv 2^N$ and they can be identified by a label 
$k$ in the interval $1-M$.

Generally speaking the Hamiltonian of the system is given when we know the 
values of the energies $E_k$ for each of the $M$ configurations of the system.  
Usually one writes explicit expression for the energies as function of the 
configuration; on the contrary here we assume that the values of the $E_k$ are 
random, with a probability distribution $p(E)$ which is Gaussian:
\be
p(E) \propto \exp (-{E^2 \over 2 N}).\label{PRO}
\ee

The partition function is simply given by
\begin{eqnarray}
Z_\cE=\sum_{k=1,M} \exp (-\beta E_k)=\int \rho(E) \exp (-\beta E),\\
\rho(E)\equiv\sum_{k=1,M} \de(E-E_k).
\end{eqnarray}
   
The value of the partition function and of the free energy density 
$(f_\cE=-\ln(Z_\cE)/(N\beta )$) depends on $\cE$, i.e.  all the values of the 
energies $E_k$.  We would like to prove that when $N
\to \infty$ the dependance on $\cE$  of $f_\cE$ disappears with probability 
1.  If this happens, the most likely value of $f_\cE$ coincides with the average 
of $f_\cE$, where the average is done respect to all the possible values of the 
energy extracted with the probability distribution \form{PRO} .

The model is enough simple to be studied in great details; exact expressions can 
be derived also for finite $N$.  Here we sketch the results giving a 
plausibility argument.

\subsection{Equilibrium properties of the model}

The crucial observation is the following.  The probability of finding a 
configuration of energy $E$ is
\be
\N_0(E)\equiv 2^N \exp(-{E^2\over 2 N}) = \exp (N(\ln(2)- \frac12 e^2)),
\ee
where $e\equiv E/N$ is the energy density.  It is reasonable to assume (and it 
is confirmed by a detailed computation) that in the case of a generic system 
(with probability 1 when $N \to \infty$) no configurations are present in the 
region where $\N_0(E)<<1$, i.e.  for
\be e^2<e_c^2\equiv 2 \ln(2).  \ee 
We can thus 
write for the generic choice of the energies ($\cE$): 
\be \rho(E)\approx 
\N(E)\equiv N_0(E)\theta(E_c^2-E^2).  
\ee
 The partition function can be written 
as 
\be \int \N(E) \exp (-\beta E) .
\ee 
Evaluating the integral with the saddle 
point method one finds that in the high temperature region, i.e.  
\be 
\beta<\beta_c\equiv e_c^{-1}, 
\ee 
the internal energy density is simple given by $-\beta$.  This behaviour must 
end somewhere because we know that the energy is bounded form below also when 
$\beta \to \infty$.  Indeed in the low temperature region one finds that the 
integral is dominated by the boundary region $E\approx E_c$ and the energy 
density is exactly given by $-e_c$.

The entropy density is positive in the high temperature region, vanishes at 
$\beta_c$ and remains zero in the low temperature region.  It follows that in 
the high temperature region an exponentially large number of configurations 
contributes to the partition function, while in the low temperature region it is 
possible that the probability is concentrated on a finite number of 
configurations.

\subsection{Properties of the low temperature phase}

It is worthwhile to study in more details the structure of the configurations 
which mostly contribute to the partition function in the lower temperature 
phase.  At this end it is useful to sort the configurations with ascending 
energy.  We rename the configurations and we introduce new labels such that 
($E_k<E_i$ for $k<i$).

It is convenient to introduce the following quantity:
\be
w_k\equiv {\exp(-\beta E_k) \over Z}.
\ee
We have obviously that
\be
\sum_{k=1,2^N}w_k =1
\ee
A detailed computation shows \cite{REM,mpv,parisibook2} that in the low 
temperature region (i.e.  $\beta>\beta_c$) the previous sum is dominated by the 
first terms.  Indeed
\be
\sum_{k=1,L}  w_k =1-O(L^{-\la}), \ \ \ \la={1-m \over m},
\ee
where 
\be
m={T \over T_c}.
\ee
In the same region the sum in the following equation is convergent and its 
average value is given by
\be
\sum _{k=1,2^N}w_k^2=1-m.
\ee

Generally speaking one finds that one can introduce the quantities $F_k$ 
such that
\be
w_k \propto \exp(-\beta F_k)
\ee
Here the variables $F_k$ coincide with the total energy (not the energy 
density!) apart form an addictive constant. 
Their
probability distribution at the lower  end (which is the relevant region for 
thermodynamics in the low temperature region) can be approximated as
\be
P(F)\approx \exp (\beta m F).
\ee

In this model everything is clear: in the high temperature region the number of 
relevant configurations is infinite (as usual) and there is a transition to a 
low temperature region where only few configuration dominates. 

This phenomenon can be seen also in the following way.  We introduce a 
distance among two configurations $\al$ and $\ga$ as
\be
d^2(\al,\ga) \equiv {\sum_{i=1,N} (\si^\al_i -\si^\ga_i)^2 \over 2 
n}.\label{DISTANZA}
\ee
Sometimes it is convenient to introduce also the overlap $q$defined as
\be
q(\al,\ga)\equiv {\sum_{i=1,N} \si^\al_i \si^\ga_i \over 2 N}=1-d^2(\al,\ga).
\ee
The distance squared is normalized in such a way that it spans the interval 
$0-2$. It is equal to
\begin{itemize}
\item
 0, if the two configuration are equal ($q=1$).
 \item 
 1, if the configuration are orthogonal ($q=0$).
 \item
 2, if $\si^\al_i=-\si^\ga_i$ ($q=-1$).
 \end{itemize}
 
It is convenient to introduce the function $Q(d)$ and $P(q)$, i.e.  the 
probability that two equilibrium configurations are at distance $d$ or 
overlap $q$ respectively.  We find 
\begin{itemize} 
\item For $T>T_c$ 
\be 
Q(d)=\delta(d-1), \ \ \ P(d)=\de(q).  
\ee 
\item For $T<T_c$ 
\be Q(d)=(1-A) 
\delta(d-1)+A\delta(d), \ \ \ P(d)=(1-A) \delta(d)+A\delta(q-1).  
\ee
where $A$is equal to $\sum _{k=1,2^N}w_k^2$.  The average of $A$ over the 
different realizations of system is equal to $1-m$.
\end{itemize}

 As soon as we enter in the low temperature region, the 
probability of finding two equal configurations is not zero.  The 
transition is quite strange from the thermodynamic point of view.
\begin{itemize}
\item
It looks like a {\sl second} order transition because there is no latent heat.
It is characterized by a jump in the specific heat (which decreases going toward 
low temperatures).
\item
It looks like a {\sl first} order transition.  There are no divergent 
susceptibilities coming form above of below (which beyond mean field theory 
should imply no divergent correlation length).  Moreover the minimum value 
of $d$ jumps discontinuously (from 1 to 0).

\item If we consider a system composed by two replicas ($\si^1$ and $\si^2$)  
\cite{KPVI} and 
we write the Hamiltonian
\be
H(\si^1,\si^2)=H(\si^1)+H(\si^2)+N \eps d^2(\si_1,\si^2)
\ee
the thermodynamics is equal to that of the previous model (apart a factor 
2) for $\eps=0$, but we find a real first order thermodynamic transition, 
with a discontinuity in the internal energy, as soon as $\eps>0$.  The case 
$\eps=0$ is thus the limiting case of real first order transitions.
\end{itemize}
 
These strange characteristics can be summarized by saying that the 
transition is of order one and half, because it share some characteristics 
with both the first order and the second order transitions.
 
It impressive to note that the thermodynamic behaviour of real glasses near 
$T_c$ is very similar to the order one and half transition of REM. We will sea 
later that this behaviour is typical of the mean field approximation to glassy 
systems.

\subsection{Dynamical properties of the model}
The dynamical properties of the model can be easily investigated in a 
qualitative way (for more rigorous results sea \cite{ISOPI} ). Interesting 
behavior is present in the region where the value of $N$ is large with respect 
to the time. Different results will be obtained for different definition of the 
dynamics if we consider some rather artificial form of the dynamics.

Let us first consider a single spin flip dynamics.  In other words we assume 
that in a microscopic time scale scale, which for simplicity we consider of 
order unit, the system explore all the configurations which differ from the 
original one by a single spin flip and goes to one of them (or remain in the 
original one) with probability which is proportional to $\exp (-\beta H)$.  This 
behaviour is typical of many dynamical process, like Glauber dynamics, monte 
Carlo, heath bath.

In this dynamical process each configuration $C$ has $N$ nearby configurations 
$C'$ to explore.  The energies of the configurations $C'$ are uncorrelated to 
the energy of $C$, so that they would be of order one in most of the case.  The 
lowest energy of the configurations $C'$ would be of order $-(N\ln(N))^{1/2}$, 
the corresponding energy density ($-(\ln(N)/N)^{1/2}$) vanishes in the large $N$ 
limit.

If the configuration $C$ has an energy density $e$ less that zero, but 
greater than the equilibrium energy, the time needed to do a transition to 
a configuration of lower energy will be, with probability one,
exponentially large.

 For 
short times a configuration of energy $e$ will be completely frozen.  Only 
at  larger times it may jump to a 
typical configuration of energy zero.  At later times different scenarios 
are possible: the configuration comes back to the original configuration of 
of energy $e$ or, after some wandering int the region of configurations of 
energy density $\approx 0$, it fells in an other deep configuration of 
energy $e'$.  A computation of the probabilities for these different 
possibilities has not yet been done, although it should not too difficult.

The conclusions of this analysis are quite simple. 
\begin{itemize}
\item
If we start from a random configuration, after a time which is finite when 
$N\to \infty$, the system goes to a configuration whose energy is of order 
$-\ln(N)^{1/2}$ and stops there.
\item If we start from a random configuration and we study the system at
exponentially large times the system will reach an energy density which may 
be different from the orginal one.
\end{itemize}

Similar results may be obtained if we consider a rather artificial dynamics in 
which we assume that in a microscopic time the 
system explore a number of configurations $V$ which is order $\exp(N h)$. The 
previous argument will tell us that the number of configurations, for which a 
transition is possible in times of order unit, is given by
\be
\exp(N(h +s(e)-s(0))).
\ee
where $s(e)=e^2/2$ is the entropy density as function of the energy density. 
The energy density which can be reached is therefore given by
\be
{e^2 \over 2} =h.
\ee

Also this case the system freezes at non zero energy. The single spin flip limit 
correspond to $h=\ln(N)/N$.

\section{Models with partially correlated energy}
\subsection{The definition of the models}
The random energy model (REM) is rather unrealistic in that it predicts that the 
energy is completely upset by a single spin flip.  This feature can be 
eliminated by considering a more refined model, the so called $p$-spins models 
\cite{GROMEZ,GARDNER} , in which the energies of nearby configurations are also 
nearby.  We could say that energy density (as function of the configurations) 
is not a continuous function in the REM, while it is continuous in the $p$-spins 
models, in the topology induced by the distance (\form{DISTANZA} ).  In this new 
case some of the essential properties of the REM are valid, but new features are 
present.

The Hamiltonian we consider depends on some control variables $J$, which have a 
Gaussian distribution and play the same role of the random energies of the REM 
and by the spin variable $\si$. For $p=1,2,3$ the Hamiltonian is respectively
\begin{eqnarray}
H^1_J(\si)= \sum_{i=1,N} J_i \si_i\\
H^2_J(\si)= \sum_{i,k=1,N}' J_{i,k} \si_i \si_k\\
H^3_J(\si)= \sum_{i,k,l=1,N}' J_{i,k,l} \si_i \si_k \si_l \nonumber
\end{eqnarray}
where the primed sum indicates that all the indices are different. The variables
$J$ must have a variance of $O(N^{(1-p)/2})$ in order to have a non trivial 
thermodynamical limit.

It is possible to prove by am explicit computation that if we send first $N \to 
\infty$ and later $p \to \infty$, one recover the REM.  Indeed the energy 
differences corresponding to one spin flip are of order $p$ for large $p$ 
(they ar order $N$ in the $REM$, so that in the limit $p \to \infty$ the 
energies in nearby configurations become uncorrelated and the REM is 
recovered.

\subsection{Equilibrium properties of the models}

The main new property of the model is the correlation of energies.  This fact 
implies that if $C$ is a typical equilibrium configuration, all the 
configurations which differ from it by a finite number of spin flips will also 
have a finite energy.  The equilibrium configurations are no more isolated (as 
in REM), but they belongs to valleys, such that the entropy restricted to a 
single valley is proportional to $N$ and it is and extensive quantity.

The thermodynamical properties at equilibrium can be computed using the replica 
method \cite{mpv,parisibook2} .  In the simplest version of this method 
\cite{GROMEZ,GARDNER} one introduces the typical overlap of two configurations 
inside the same valley (sometimes denoted by $q_{EA}$.  Something must be said 
about the distribution of the valleys.  Only those which have minimum free 
energy are relevant for the thermodynamics.  One finds that these valleys have 
zero overlap and have the same distribution of free energy as in the REM
\be
P(F)\propto \exp(\beta m (F-F_0)).
\ee
Indeed the average value of the free energy can be written in a self consistent 
way as function of $m$ and $q$ ($f(q,m)$) and the value of these two parameters 
can be found as the solution of the stationarity equations:
\be
\parziale{f}{m} =\parziale{f}{q} =0.
\ee

The quantity $q$ (which would be 1 in the REM) is here of order $1-\exp(-A\beta 
p)$ for large $p$, while the parameter $m$ has the same dependence of the 
temperature as in the REM, i.e.  1 at the critical temperature, and a linear 
behaviour al low temperature.  The only difference is that $m$ is no more 
strictly linear as function of the temperature.

The thermodynamical properties of the model are the same as is the REM: a 
discontinuity in the specific heat, with no divergent susceptibilities.
\subsection{The Free energy landscape}
It would be interesting to characterize better  the free energy landscape of 
the model, especially in order to understand the dynamics. Indeed we have 
already seen that in the REM the system could be trapped in metastable 
configurations. Here the situation is more complicated. Although the word {\sl 
valley} has a strong intuitive appeal, we must first define 
what a valley is in a more precise way.

There are two different (hopefully equivalent) definitions of a valley:
\begin{itemize}
\item
A valley is a region of configuration space separated by the rest of the 
configuration space by free energy barriers which diverge when $N\to\infty$.  
More precisely the system, in order to go outside a valley by moving one spin at 
once, must cross a region where the free energy is higher that of the valley by 
a factor which goes to infinity with $N$.
\item
A valley is a region of configuration space in which the system remains for time 
which goes to infinity with $N$.
\end{itemize}
The rationale for assuming that the two definitions are equivalent is the 
following.  We expect that for any reasonable dynamics in which the systems 
evolves in a continuous way (i.e.  one spin flip at time), when it goes from a 
valley to an other valley, the system must cross a configuration of higher free energy 
and therefore the time for escape from a valley is given by
\be
\tau \simeq \tau_0 \exp (\beta \Delta F)
\ee
where $\Delta F$ is the free energy barrier.

It is crucial to realize that in  infinite range models there can be valley which 
have an energy density higher that that of equilibrium states. This phenomenon 
is definitely not present in short range models. No metastable states with 
infinite mean life do exist in nature. 

Indeed let us suppose that the system may stay in phase (or valleys) which we 
denote as $A$ and $B$.  If the free energy density of $B$ is higher than that of 
$A$, the system can go from $B$ to $A$ in a progressive way, by forming a bubble 
of radius $R$ of phase $A$ inside phase $B$.  If the surface tension among phase 
$A$ and $B$ is finite, has happens in any short range model, for large $R$ the 
volume term will dominate the free energy difference among the pure phase $B$ 
and phase $B$ with a bubble of $A$ of radius $R$.  This difference is thus 
negative at large $R$, it maximum will thus be finite.  In the nutshell a finite 
amount of free energy in needed in order to form a  seed of phase $A$ starting 
from which the spontaneous formation of phase $A$ will start. For example, if 
we take a mixture of $H_2$ and $O_2$ at room temperature, the probability of a 
spontaneous temperature fluctuation  in a small region of the sample, 
which lead to later ignition and eventually to the explosion of the whole 
sample, is greater than zero (albeit quite a small number), and obviously it does not go 
to zero when the volume goes to infinity.

We have two possibilities open in positioning this mean field theory prediction 
of existence of real metastable states:
\begin{itemize}
	\item  We consider the presence of these metastable state with {\sl infinite} 
	mean life an artefact of the mean field approximation and we do not pay 
	attention to them.

\item We notice that in the real systems there are metastable states with very 
large (e.g.  much greater than one year) mean life.  We consider the {\sl 
infinite} time metastable states of the mean field approximation as precursors 
of these {\sl finite} states.  We hope (with reasons) that the corrections to 
the mean field approximation will give a finite (but large) mean life to 
these states (how this can happen will be discussed in the next section).
\end{itemize}

Here we suppose that the second possibility is the most interesting and we 
proceed with the study of the system in the mean field approximation.  The 
strategy for investigate the properties of these metastable states consists in 
considering systems with $R$ replicas (two or more) of the same system with 
Hamiltonian given by
\be
\beta H = \sum_{r=1,R}\beta_r H(\si^r) + \sum_{r,s=1,R}\eps_{r,s} q_{r,s}.
\ee
Different replicas may stay at different temperature. The quantities $\eps$ are 
just Legendre multipliers needed to enforce  specific value of the the overlaps 
$q$. In this way (let us consider for simplicity the case where all temperature 
are zero) we find (after a Legendre transform) a free energy density as function 
of the $q$. 

Let us consider for simplicity the case where we set
\be
q_{1,r}= q \for r=2, R
\ee
($q_{r,r}$ is identical equal to 1) and the others $q$ are left free \cite 
{KPVI} .
A simple computation show the final free energy density  that we obtain (let us call 
$f_R(q)$ is given by
\begin{eqnarray}
f_R(q) - R f= - \lim_{N \to \infty} {\ln
 \( \sum_\si \( \sum_\tau \delta(q(\si,\tau)-q) \cp ^{(R-1)} \cp
 \over
 \beta N}\equiv\\
- \lim_{N \to \infty} {\ln <P_\si(q)^{(R-1)}> \over \beta N},
\end{eqnarray}
where $f$ is the unconstrained free energy.

A particular case, which is very interesting, is given by the limit $R \to 1$ 
\cite{KLEIN,FP} :
\be
W(q)\equiv \lim_{N \to \infty} {\ln (P_\si(q))\over \beta N}=
\parziale{(f_R(q) - R f)}{R} |_{R=1}
 \ee

The potential $W(q)$ has usually a minimum at $q=0$, where $W(0)=0$. It may have a secondary 
minimum at $q=q_D$. We can have three quite different situations
\begin{itemize}
\item $W(q_D)=0$.  This happens in the low temperature region, below $T_c$, 
where we can put two replicas both at overlap $0$ and at overlap $q_{EA}$ 
without paying any prize in free energy.  In this case $q_D=q_{EA}$.

\item $W(q_D)>0$.  This happens in an intermediate temperature region, above 
$T_c$, but below $T_D$, where we can put one replica $\si$ at equilibrium and 
have the second replica $\tau$ in a valley near it. It happens that the 
internal energy of both the $\si$ configuration (by construction) and of the 
$\tau$ configuration are equal to the equilibrium one. However the number of 
valley is exponentially large so that the free energy a single  valley will be 
smaller. One finds in this way that $W(q_D)>0$ is given by
\be
W(q_D)= {\ln \N _e \over N}
\ee
where $ \N _e $ is the average number of the 
valleys having the equilibrium energy \cite{MONA,PP} .

	\item  At $T>T_D$ the potential $W(q)$ has only the minimum at $q=0$. The 
	quantity $q_D$ cannot be defined and no valley 
	with the equilibrium energy are present. This is more or less the definition 
	of the dynamical transition temperature $T_D$. A more careful analysis shows that 
	for $T_D<T<T_V$ there are still valleys with energy {\sl less} than the equilibrium 
	one, but these valleys cover a so small region of phase space that they are not 
	relevant for equilibrium physics.
\end{itemize}

It is also possible to study the properties of the free energy for $R\ne 1$ we 
can force the $\si$ configuration to be not an equilibrium one and in this way 
we controll the properties of  the valleys having an energy different than the 
equilibrium one.
This method can give rather detailed information on the free energy landscape, 
which I do not have time to discuss in details and which have not yet yet fully 
studied.

The most interesting result is that for $T<T_D$ the entropy of the system can be 
written as
\be
S= S_V + W
\ee
where $S_V$ is the entropy inside a valley and $W$ is the configurational 
entropy, or complexity, i.e. the term due to the existence of an exponentially 
large number of states. The $W$ contribution vanishes at $T_c$ and becomes 
exactly equal to zero for $T<T_c$ \cite{KT1,KT2,KT3} .

In the REM limit ($p\to \infty$) the temperature $T_D$ goes to infinity.  
In this limit the third region does not exist.  Therefore the dynamical 
transition is a new feature which is not present in the REM.

\subsection{Mode coupling theory at short times}

The simplest way to study the dynamics of the problem is to consider a 
system which evolves according to Langevin equation of to some sort of 
Glauber dynamics.  For example we can suppose that:
\be
{d \si_i \over dt} = -{\de H \over \si_i} +\eta_i(t),
\ee
where $\eta$ is an appropriate white noise.

In this model is convenient to introduce the
single site
correlation function ($C$) and the response ($G$) function of the times.
One finds that they can be defined as 
\begin{eqnarray}
C(t_1,t_2)=<\si_i(t_1) \si_i(t_2)>\\
G(t_1,t_2)=<{\de \si_i(t_1) \over h_i(t_2)}>
\end{eqnarray}
where $h(t)$ is an external magnetic field. 

If the systems is at equilibrium (or in a metastable state), the 
correlation  and the response functions will depend only on the time 
difference. If the system is out of equilibrium these functions will 
depend in a non trivial way from both the arguments.

In both cases one can write down closed equations for the correlation 
functions in the case where the size $N$ goes to infinity at fixed times 
\cite{CUKU} .  These 
equations have a rather complex structure.  We could discuss two different 
regimes:
\begin{enumerate}
	\item  We start at time zero from an equilibrium configuration.

	\item  We  start at time zero from a random configuration.

\end{enumerate}
In the first case we find that, if we approach the dynamical temperature 
from above, the correlation time diverges a a power of $T-T_D$, and the 
usual analysis of the mode coupling theory can be done in this region. Mode 
coupling theory is essentially correct in this region.

In the second case one finds that below the dynamical transition, the energy 
does not go anymore to the equilibrium value, but it goes to an higher value, 
which in some cases can be computed analytically.

The phenomelogy is rather complex, the aging properties of the systems are 
particularly interesting, but we cannot discuss them for lack of space. It is 
interesting to note that the mode coupling theory become exact in the mean 
field theory and describes what happens nearby the dynamical phase transition 
at $T_D$, which is a temperature  higher that the equilibrium transition 
temperature at $T_c$.

\subsection{Dynamical properties of the models from a landscape analysis}

Here we would like to understand the dynamical properties of the model starting 
from results which we have already derived on the free energy landscape without 
having to compute explicitly the solution of the equation of motion for the 
correlations $C$ and $G$. This can be done by assuming the this very slow 
dynamics is an activated process dominated by barrier crossing and that the 
height of the barriers can be computed using the method of the previous 
sections. 

Some of the question we ask are the following:
\begin{itemize}
\item How long does a system remains  in the same 
valley?
\item If the system is out equilibrium at time 0, which is the asymptotic value of 
the energy at large times?
\end{itemize}
 The results are 
similar to those obtained by the explicit dynamical analysis of the previous 
section.  We obtain also more detailed information on the region where the time 
scale is exponentially large.  This region can be studied only with very great 
difficulties using the dynamical equations.

In the first case (i.e.the system start from a thermalized configuration)
below $T_D$, the system is confined to a valley.
A first estimate of the time needed to escape from a valley can be obtained as 
follows. We introduce the parameter 
\be
q(t)\equiv {\sum_{i=1,N} \si_i(0) \si_i(t) \over N}.
\ee

We assume that on the large time scale the evolution of $q$ is more or less the 
same of a system with only one degree of freedom with potential $W(q)$. This 
assumption is not completely correct, but it is likely to be enough to give a 
first estimate (to be refined later) of the escape time.
Therefore it is reasonable to assume that in a first approximation
the the time needed to escape from the valley is given by 
\be
\tau=\tau_0 \exp(N \beta \De W),
\ee
where $\De W$ is the difference in free energy among the minimum at $q_D$ 
and the maximum al lower values of $q$.

For large time (but not exponentially divergent with $N$) we have that
\be
\lim_{t \to \infty}q(t) = q_D.
\ee
Eventually the system will escape from a valley for times greater that 
$\tau$.

In the other situation (the system is originally out equilibrium) things are 
more complicated.  Metastable valleys of high energies do exist, however it is 
not clear that a system cooled from an high temperature region must be trapped 
in the valley with high energy.  Indeed it will be trapped in the valley with 
largest attraction domain, but the properties of the valley cannot be found if 
we do not use the dynamics in an explicit way.  Some educated conjectures can be 
done, but the question has never been investigated in details.  The problem is 
not easy, because the system is strongly out of equilibrium.

A better understanding comes if we consider the case in which the system is 
slowly cooled, from high to low temperature. We consider here that case of 
slow, not ultra slow, cooling, i.e. the time scale if fixed when $N$ goes 
to infinity. In this case it is reasonable to assume that the system frozen 
in a valley at the dynamical temperature and we have to follow the energy 
of that valley when we cool. That can be done by using the formulation with 
different replicas at different temperatures. In this way one finds that 
the system is frozen in a configuration which is near to the configuration 
at the dynamical temperature.

A detailed comparison of the results for the energy of the metastable states 
obtained by doing different assumption is 
still lacking, but is should be not too difficult.

\section{More realistic models}
\subsection{General consideration on long range Hamiltonians with and without quenched 
disorder}

 A peculiar phenomenon, that is present in Hamiltonians when the range 
is infinite, is the partial equivalence of Hamiltonians with quenched and random 
disorder.  An example is the following.

The configuration space is given by $N$ Ising spin variables. We consider the 
following Hamiltonian
\begin{eqnarray}
\sum_{i=1,N}|B_i|^2-1|^2, \\
B_i=\sum_{k=1,N} R_{i,k} \si_k.
\end{eqnarray}
where $R$ is an unitary matrix, i.e.
\be
\sum_{k=1,N}R_{i,k}\ba{R_{k,m}}=\de_{i,m}.
\ee

We could consider two different cases \cite{MPR} :
\begin{itemize}
\item The matrix $R$ is a random orthogonal matrix.
\item The matrix $R$ is given by
\be
R(k,m) ={\exp (2 \pi i \ k m) \over N^{1/2}}
\ee
In other words $B$ is the Fourier transform of $\si$.
\end{itemize}

The second case is a particular instance of the first one, exactly in the 
same way that a sequence of all zeros is a particular instance of a random 
sequence.

The first model can be studied using the replica method and one finds results 
very similar to those of the $p$-spin model we have already studied.

Now it can be proven that the statistical properties of the second model ar 
identical to those of the first model, with however an extra phase.  In the 
second model (at least for some peculiar value of $N$, e.g.  $N$ prime, 
\cite{MIGLIO,MPR,BGU} ) there are configurations which have exactly zero energy.  These 
configuration form isolated valleys which are separated from the others, but 
have much smaller energy and they have a very regular structure (like a 
crystal).  An example of these configurations is
\be
\si_k \equiv_{ \bmod  N }k^{(N-1)/2}
\ee
(The property $k^{(N-1)}\equiv 1$ for prime N, implies that in the previous 
equations $\si_k=\pm 1$).  Although the sequence $si_k$ given by the previous 
equation is apparent random, it satisfies so many identities that it must be 
considered as an extremely ordered sequence (like a crystal).  One finds out that from the 
thermodynamical point of view it is convenient to the system to jump to one of 
these ordered configurations at low temperature.  More precisely there is a first 
order transition (like a real crystalization transition) at a temperature, which 
is higher that the dynamical one.

If the crystallisation transition is avoided by one of the usual methods, 
explicit interdiction or sufficient slow cooling, the properties of the second 
model are exactly the same of those of the first model.  Similar considerations 
are also valid for other spin models \cite{Frhe,MPRII,MPRIII} or for model of 
interacting particles in very large dimensions, where the effective range of the 
force goes to infinity \cite{CKPR,CKMP,PR} .

  We have seen that by removing 
the quenched disorder in the Hamiltonian had a quite positive effect: a 
crystallisation transition is also present like in some real systems.  If we 
neglect crystalization, which is absent for some values of $N$, no new feature 
is present in system without quenched disorder. As we shall see later the 
equivalence of short range systems with and without quenched disorder is an 
interesting and quite open problem.

\subsection{Short range models}

We have already seen that in a short range model we cannot have real metastable 
states.  Let us see in more details what happens.

Let us assume that $\De f<0$ is the difference in free energy among the 
metastable state and the stable state. 
Now let us consider a bubble of radius $R$ of stable state inside the metastable 
one. The free energy difference of such a bubble will be
\be
F(R) =-\De f \ V(R) - I(R)
\ee
where the interfacial free energy $I(R)$ can increase at worse as $\si
\Sigma(R)$. the quantities $V(R) \propto R^D$ and $\Sigma(R) \propto R^{D-1}$ are the volume 
and the surface of the bubble and $\si$ is the surface tension, which can also 
be zero (when the surface tension is zero, we have $I(R)\propto R^\omega$, with 
$\omega<D-1$).
 
The value of $F(R)$ increases at small $R$, reaches a maximum a $R_c$, which in 
the case $\si\ne 0$ is of order $(\De f) ^{-1}$ and it becomes eventually 
negative at large $R$.  According to enucleation theory, the system goes from 
the metastable to the stable phase under the formation and the growth of such 
bubbles, and the time to form one of them is of order (neglecting prefactors)
\begin{eqnarray}
\ln (\tau) \propto \De f\  R^D \propto (\De f)^{(D-1)}\\
\tau \propto \exp( {A \over (\De f)^{(D-1)}})
\end{eqnarray}
where $A$ is constant dependent on the surface tension.
In the case of zero surface tension we have
\begin{eqnarray}
\tau \propto \exp( {A \over \De f^\la})\\
\la={D \over \omega} -1
\end{eqnarray}

This argument for the non existence of metastable states can be naively applied 
here.  The metastable states of the the mean field approximation now do decay.  
The dynamical transition becomes a smooth region which separates different 
regimes; an higher temperature regime where mode coupling theory can be 
approximately used and a low temperature region where the dynamics is dominated 
by barriers crossing.  There is no region where the mode coupling theory becomes 
exact but it is only an approximated theory which describe the behaviour in a 
limited region of relaxation times (large, but not too large).

The only place where the correlation time may diverge is at that the 
thermodynamical transition $T_c$, whose existence seems to be a robust 
prediction of the mean field theory.  It follows that the only transition, both 
from the static and the dynamical point of view, is present at $T_c$.

In order to understand better what happens near $T_c$ we must proceed in a 
careful matter.  The nucleation phenomenon which is responsible of the decay of 
metastable states is of the same order of other 
non-perturbative corrections to the mean field behaviour, which cannot be seen 
in perturbation theory.  We must therefore compute in a systematic way all 
possible sources of non perturbative corrections.  This has not yet been done, 
but it should not be out of reach. 

One of the first problem to investigate is the equivalence of systems with and 
without random disorder. In systems with quenched disorder there are 
local inhomogeneities which correspond to local fluctuations of the critical 
temperature and may dominate the thermodynamics when we approach the critical 
temperature \cite{FPII} . 

It is quite possible, that systems with and without quenched disorder, although 
they coincide in the mean field approximation, they will be quite different in 
finite dimension (e.g.  3) and have different critical exponents.  The scope of 
the universality classes would be one of the first property to assess.

It is not clear at the present moment if the 
strange one order and half transition is still present in short range model or 
if it is promoted to a bona fide second order transition. If the transition 
remains of the order one and half (for example is conceivable that this 
happens only for systems without quenched disorder). It could also 
possible that there is appropriate version of the enucleation theory which is 
valid near $T_c$ and  predicts:

\be
\tau\propto \exp( {A \over (T-T_C)^\al}).
\ee

A first guess for $\al$ is $D-2$ \cite{KT3, KLEIN} , although other values, e.g.  
2/3, are possible. A more detailed understanding of the static properties near 
$T_c$ is needed before we can do any reliable prediction.

It seems to me that the theoretical situation is quite good: we do not have yet 
the answers to our questions, but is seems that we are starting to ask the right 
questions. There is a lot of hard work waiting for us.

\end{document}